\documentclass[11pt, a4paper]{article}
\pdfoutput=1
\usepackage{jcappub}

\usepackage{booktabs}
\usepackage{multirow}
\usepackage{amssymb}
\usepackage[export]{adjustbox}
\usepackage{floatrow}
\newfloatcommand{capbtabbox}{table}[][3.5in]
\newfloatcommand{capbfigbox}{figure}[][2.3in]
\usepackage{blindtext}
\usepackage{amsmath}
\usepackage{booktabs}
\usepackage{aas_macros}

\def\lsim{\mathrel{\raise.3ex\hbox{$<$\kern-.75em\lower1ex\hbox{$\sim$}}}}
\def\gsim{\mathrel{\raise.3ex\hbox{$>$\kern-.75em\lower1ex\hbox{$\sim$}}}}

\begin{document}

\hspace*{110mm}{\large \tt FERMILAB-PUB-17-173-A}

\vskip 0.2in

\title{TeV Gamma Rays From Galactic Center Pulsars}

\author{Dan Hooper,$^{a,b,c}$}\note{ORCID: http://orcid.org/0000-0001-8837-4127}
\emailAdd{dhooper@fnal.gov}
\author{Ilias Cholis$^{d}$}
\emailAdd{icholis1@jhu.edu}\note{ORCID: http://orcid.org/0000-0002-3805-6478}
\author{and Tim Linden$^e$}\note{ORCID: http://orcid.org/0000-0001-9888-0971}
\emailAdd{linden.70@osu.edu}

\affiliation[a]{Fermi National Accelerator Laboratory, Center for Particle Astrophysics, Batavia, IL 60510}
\affiliation[b]{University of Chicago, Department of Astronomy and Astrophysics, Chicago, IL 60637}
\affiliation[c]{University of Chicago, Kavli Institute for Cosmological Physics, Chicago, IL 60637}
\affiliation[d]{Department of Physics and Astronomy, The Johns Hopkins University, Baltimore, Maryland, 21218}
\affiliation[e]{Ohio State University, Center for Cosmology and AstroParticle Physics (CCAPP), Columbus, OH  43210}

\abstract{Measurements of the nearby pulsars Geminga and B0656+14 by the HAWC and Milagro telescopes have revealed the presence of bright TeV-emitting halos surrounding these objects. If young and middle-aged pulsars near the Galactic Center transfer a similar fraction of their energy into TeV photons, then these sources could plausibly dominate the emission that is observed by HESS and other ground-based telescopes from the innermost $\sim$$10^2$ parsecs of the Milky Way. In particular, both the spectral shape and the angular extent of this emission is consistent with TeV halos produced by a population of pulsars, although the reported correlation of this emission with the distribution of molecular gas suggests that diffuse hadronic processes also must contribute. The overall flux of this emission requires a birth rate of $\sim$100-1000 neutron stars per Myr near the Galactic Center, in good agreement with recent estimates.}

\maketitle

\section{Introduction}

The HESS, VERITAS and MAGIC Collaborations have each reported the detection of very high-energy (VHE) gamma-ray emission from the direction of the Galactic Center, extending to energies of $\sim$30-50 TeV~\cite{Abramowski:2016mir,Ahnen:2016ujd,Aharonian:2004wa,Kosack:2004ri,Albert:2005kh,Aharonian:2009zk,Archer:2014jka,Collaboration:2009tm}. When this emission was initially identified, it was suggested that it may originate from the Milky Way's central supermassive black hole, Sgr A$^*$~\cite{Aharonian:2004jr,Atoyan:2004ix,Chernyakova:2011zz,Linden:2012iv}. More recent measurements, however, have revealed that this emission includes a component that is extended to $\sim$$10^2$ parsecs in radius~\cite{Abramowski:2016mir}. In light of this, it has been proposed that cosmic rays originating from Sgr A$^*$ may be responsible for the observed VHE emission. Because multi-TeV electrons would lose energy through inverse-Compton scattering and synchrotron processes too rapidly to account for the observed extension, this emission has instead been interpreted as evidence that Sgr A$^*$ accelerates cosmic-ray protons up to $\sim$PeV energies, which then propagate outward and generate the observed VHE gamma-ray emission through pion production~\cite{Abramowski:2016mir,Fujita:2016yvk,Guo:2016zjl}.

In this paper, we revisit the origin of the VHE gamma rays observed from the Inner Galaxy (not to be confused with the GeV excess observed by the Fermi Telescope~\cite{Goodenough:2009gk,Hooper:2010mq,Hooper:2011ti,Abazajian:2012pn,Hooper:2013rwa,Gordon:2013vta,Daylan:2014rsa,Calore:2014xka,TheFermi-LAT:2015kwa,TheFermi-LAT:2017vmf}) and offer an alternative interpretation in terms of the inverse-Compton scattering of VHE electrons/positrons generated by a population of centrally located pulsars (see also Refs.~\cite{Kistler:2015yrf,Kistler:2015toa}). In order to calculate the intensity, spectrum and spatial morphology of TeV gamma-ray emission from pulsars in the Galactic Center, we consider the nearby and well-characterized pulsars Geminga and B0656+14 (i.e. the Monogem pulsar) and treat them as representative systems. To this end, we utilize observations of these pulsars as reported by the the HAWC~\cite{Abeysekara:2017hyn,2015arXiv150803497B,2016JPhCS.761a2034C,2015arXiv150907851P} and Milagro~\cite{2009ApJ...700L.127A} Collaborations. The angular extension observed by these telescopes strongly favor an inverse-Compton origin of this emission~\cite{Yuksel:2008rf,Hooper:2017gtd}, and the observed flux indicates that a significant fraction of the spin-down power from these pulsars is transferred into the production of VHE leptons (between 7.2\% and 29\% in the case of Geminga, across the range of models considered in Ref~\cite{Hooper:2017gtd}). This conclusion is further supported by the detections of TeV halos around young pulsars by HESS~\cite{Linden:2017vvb,Abdalla:2017vci}.

The large number of massive stars and low-mass X-ray binaries present in the Galactic Center indicates that that this region is likely to host a large population of neutron stars~\cite{Muno:2005dy}. The authors of Ref.~\cite{Pfahl:2003tf}, for example, estimate that $\sim$$10^2-10^3$ radio pulsars should be located within the innermost 0.02 parsecs around Sgr A$^*$. Intriguingly, none of these pulsars have been detected. Until recently, it had been argued that the absence of observed radio pulsars near the Galactic Center was likely due to a large free electron density in the central parsec, which creates significant dispersion in the radio pulse~\cite{1978ApJ...222L...9B,Bower:2006dp}. However, the 2013 observation of pulsations from the magnetar SGR J1745-29~\cite{Kennea:2013dfa,Mori:2013yda,Rea:2013pqa,Eatough:2013nva}, located only $\sim$$0.1$ parsecs from Sgr A$^*$, has forced a revision of this view, suggesting that there may be fewer radio pulsars in the Galactic Center than previously expected. The authors of Ref.~\cite{Dexter:2013xga}, for example, conclude that this information constitutes a ``missing pulsar problem'', and suggests that a possible resolution could be the efficient formation of magnetars (rather than ordinary pulsars). In contrast, the authors of Ref.~\cite{Chennamangalam:2013zja} argue that it is premature to conclude that the number of Galactic Center pulsars is small, and derive a conservative upper limit of $\sim$200 potentially observable pulsars located within in innermost parsec. More generally speaking, it remains widely anticipated that there are many pulsars located near the Galactic Center~\cite{Rajwade:2016cto,Chennamangalam:2013zja,Zhang:2014kva}.

In this paper, we operate under the assumption that the VHE emission from Geminga and B0656+14 is typical of that from pulsars, including those located in the Inner Galaxy. We find that the observations of the Galactic Center region by HESS and other ground-based telescopes can easily be accommodated by a population of young and middle-aged pulsars. In particular, the spectral shape and angular extent of the observed VHE emission is consistent with a population of pulsars that are born in the innermost parsec and which subsequently migrate outward as a result of pulsar natal kicks. The overall normalization of the observed emission requires a recent average birth rate of $\sim$100-1000 neutron stars per Myr near the Galactic Center.

\section{The Gamma Ray Spectrum From Electrons Around Pulsars}

High-energy electrons and positions undergo energy losses through a combination of inverse-Compton and synchrotron processes, at a rate given by~\cite{Blumenthal:1970gc}:
\begin{eqnarray}
-\frac{dE_e}{dt}(r) &=& \sum_i \frac{4}{3}\sigma_T \rho_i(r) S_i(E_e) \bigg(\frac{E_e}{m_e}\bigg)^2 + \frac{4}{3}\sigma_T \rho_{\rm mag}(r) \bigg(\frac{E_e}{m_e}\bigg)^2, 
\label{EL}
\end{eqnarray}
where $\sigma_T$ is the Thomson cross section. The quantity $S_i(E_e)$ quantifies the suppression of inverse-Compton scattering in the Klein-Nishina regime ($E_e \gsim m^2_e/2T$), and is given by:
\begin{equation}
S_i (E_e) \approx \frac{45 \, m^2_e/64 \pi^2 T^2_i}{(45 \, m^2_e/64 \pi^2 T^2_i)+(E^2_e/m^2_e)}.
\end{equation}
The sum in Eq.~\ref{EL} is carried out over the various components of the radiation backgrounds, consisting of the cosmic microwave background (CMB), infrared emission (IR), starlight (star), and ultraviolet emission (UV). In our previous study focusing on the nearby Geminga and B0656+14 pulsars~\cite{Hooper:2017gtd}, we adopted the following parameters: $\rho_{\rm CMB}=0.260$ eV/cm$^3$, $\rho_{\rm IR}=0.60$ eV/cm$^3$, $\rho_{\rm star}=0.60$ eV/cm$^3$, $\rho_{\rm UV}=0.10$ eV/cm$^3$, $\rho_{\rm mag}=0.224$ eV/cm$^3$ (corresponding to $B=3\,\mu$G), and $T_{\rm CMB} =2.7$ K, $T_{\rm IR} =20$ K, $T_{\rm star} =5000$ K and $T_{\rm UV} =$20,000 K. In the region surrounding the Galactic Center, however, we expect the energy densities of the radiation and magnetic fields to be significantly higher than those found in the local environment.

\begin{figure}
\includegraphics[width=3.08in,angle=0]{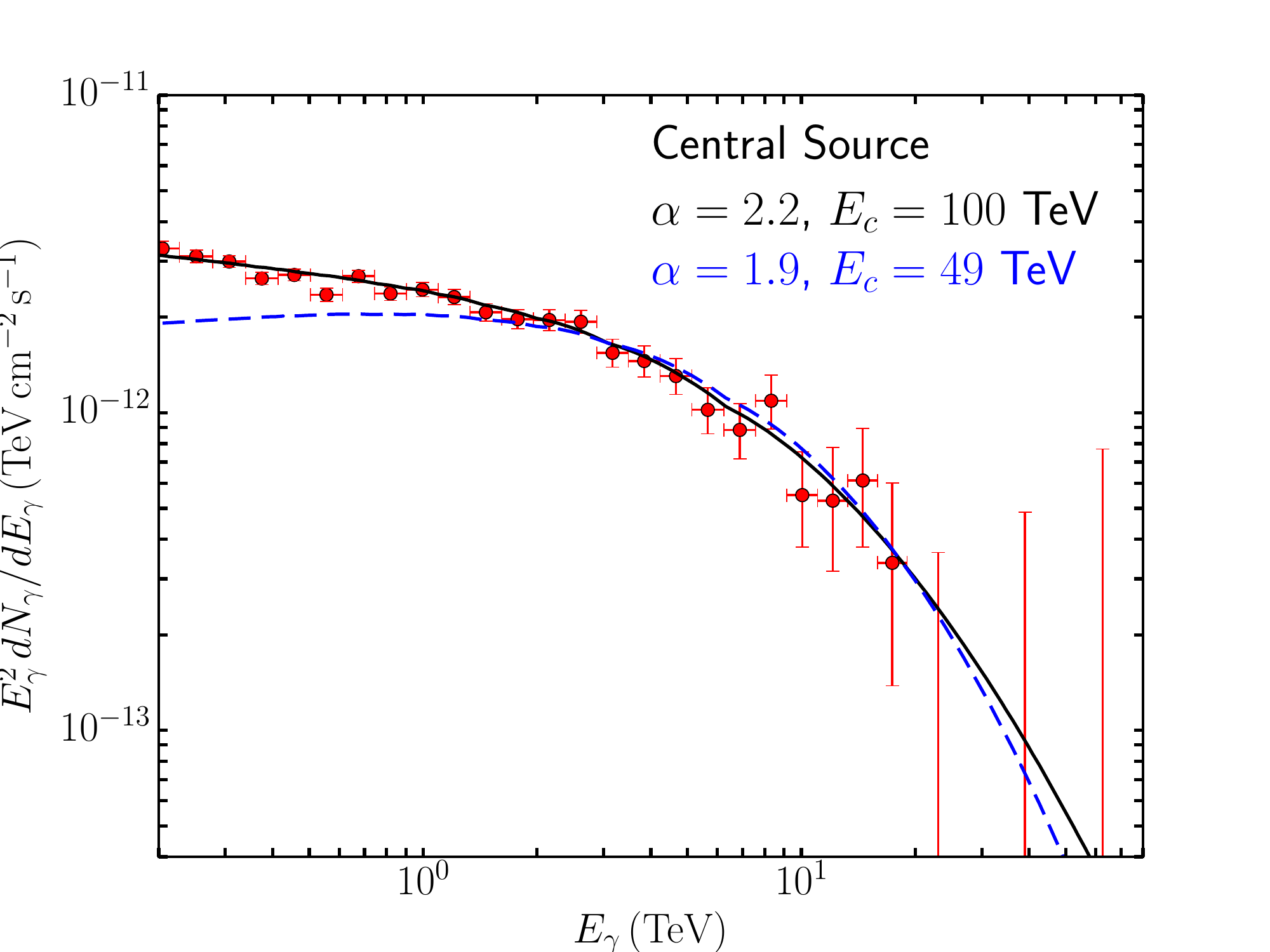}
\hspace{-0.5cm}
\includegraphics[width=3.08in,angle=0]{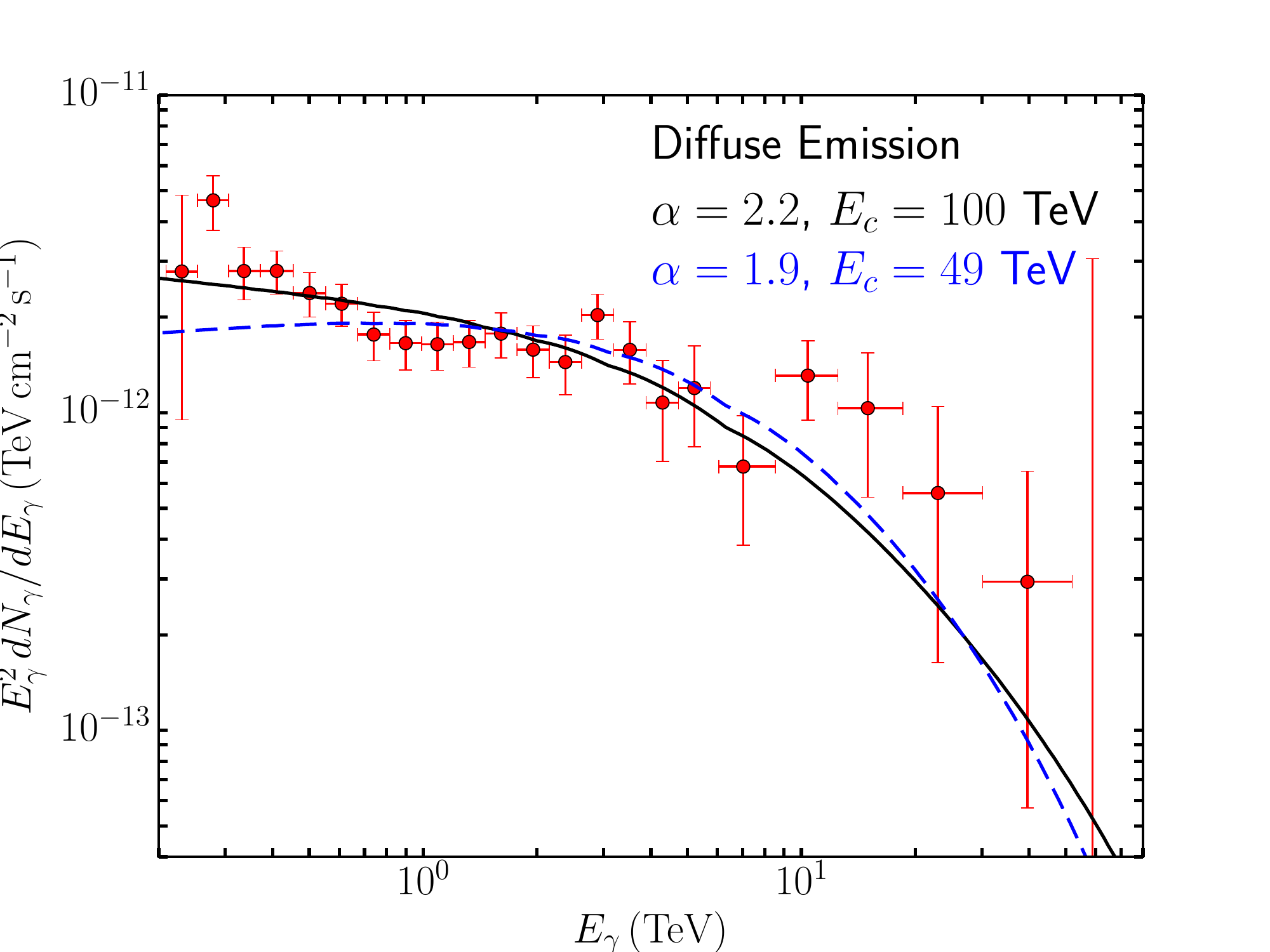}
%
\caption{The gamma-ray spectrum from the inverse-Compton scattering of very high-energy electrons and positrons in the region surrounding the Galactic Center. In the left frame, we show the spectrum of the central point source as reported by the HESS Collaboration~\cite{Abramowski:2016mir}, which is compared to that predicted in an environment with energy densities of radiation and magnetic fields that are 1000 times higher than in the local interstellar medium. In the right frame, we show the spectrum reported by HESS from within a $0.2^{\circ}$ to $0.5^{\circ}$ (partial) annulus around the Galactic Center, compared to that predicted with radiation and magnetic field energy densities that are ten times higher than in the local environment. We have parameterized the injected electron spectrum using the form $dN_e/dE_e \propto E_e^{-\alpha} \, \exp(-E_e/E_c)$, and show results in each frame for parameter values chosen to match the observed spectrum from Geminga ($\alpha=1.9$, $E_c=49$ TeV), and for values chosen to provide the best-fit to the combination of the point-source and diffuse gamma-ray observations shown ($\alpha=2.2$, $E_c=100$ TeV).}
\label{gammaspec}
\end{figure}

In Fig.~\ref{gammaspec}, we plot the gamma-ray spectrum that results from the inverse-Compton scattering of VHE electrons and positrons from pulsars in the region surrounding the Galactic Center. In each frame, we have parameterized the injected electron spectrum using the form $dN_e/dE_e \propto E_e^{-\alpha} \, \exp(-E_e/E_c)$, and show results for parameter values chosen to match the observed spectrum from Geminga ($\alpha=1.9$, $E_c=49$ TeV)~\cite{Hooper:2017gtd}, and for values chosen to provide the best-fit to the spectra shown ($\alpha=2.2$, $E_c=100$ TeV). In the left frame, we compare the predicted spectrum to that of the central point source as reported by the HESS Collaboration, while in the right frame we show the spectrum reported by HESS in a $0.2^{\circ}$ to $0.5^{\circ}$ (partial) annulus around the Galactic Center~\cite{Abramowski:2016mir}. For the central point source (extended annulus), we calculate the spectrum of inverse Compton emission assuming energy densities of starlight, IR and UV radiation and magnetic fields that are 1000 (10) times higher than in the local interstellar medium. While these energy densities represent a very approximate estimate, we consider it to be reasonable for the inner volume of the Milky Way~\cite{Genzel:2010zy,Do:2013sm,Lu:2013sn,Cholis:2014fja,Wolfire:1990zz}. The main impact of this choice is to reduce the role of scattering with the CMB, and the precise values of these quantities does not strongly impact our results or conclusions. For each curve, the overall normalization was independently chosen to to provide the best-fit to the gamma-ray spectrum reported by HESS.

\section{Modeling the Galactic Center Pulsar Population}
\label{model}

In the previous section, we demonstrated that the spectrum of the gamma-ray emission observed from the Inner Galaxy by HESS is consistent with the TeV halo emission observed from Geminga and B0656+14. We have not yet, however, discussed the normalization of the flux of VHE gamma-rays from the Galactic Center pulsar population, which depends on the evolution of these sources. 

The spin-down power of a given pulsar (the rate at which it loses rotational kinetic energy through magnetic dipole braking) is given by~\cite{Gaensler:2006ua}:
\begin{eqnarray}
\label{edot}
\dot{E} &=& -\frac{8\pi^4 B^2 R^6}{3 c^3 P(t)^4} \\
&\approx& 1.0 \times 10^{35} \, {\rm erg}/{\rm s} \times \bigg(\frac{B}{1.6 \times 10^{12} \, {\rm G}}\bigg)^2 \, \bigg(\frac{R}{15 \,{\rm km}}\bigg)^6 \, \bigg(\frac{0.23 \, {\rm s}}{P(t)}\bigg)^4, \nonumber
\end{eqnarray}
where $B$ is the strength of the magnetic field at the surface of the neutron star, $R$ is the radius of the neutron star, and the rotational period evolves as follows:
\begin{eqnarray}
P(t) = P_{_0} \, \bigg(1+\frac{t}{\tau}\bigg)^{1/2},
\end{eqnarray}
where $P_{_0}$ is the initial period, and $\tau$ is the spindown timescale:
\begin{eqnarray}
\label{tau}
\tau&=&\frac{3c^3 I P_{_0}^2}{4\pi^2B^2 R^6} \\
&\approx& 9.1 \times 10^3 \,{\rm years} \,\times \, \bigg(\frac{1.6\times 10^{12}\,{\rm G}}{B}\bigg)^2\,\bigg(\frac{M}{1.4 \, M_{\odot}}\bigg) \, \bigg(\frac{15 \, {\rm km}}{R}\bigg)^4 \, \bigg(\frac{P_{_0}}{0.040 \, {\rm sec}}\bigg)^2. \nonumber 
\end{eqnarray}

To model the population of pulsars born in and around the Galactic Center, we adopt the distribution of initial periods and magnetic fields described in Ref.~\cite{Bates:2013uma}. More specifically, for the initial period we adopt a normal distribution with $\langle P_0 \rangle =0.15$ s and  $\sigma=0.3$ s, while for the magnetic field we adopt a log-normal distribution with $\langle \log_{10} B (G) \rangle =12.65$ and  $\sigma=0.55$.

Focusing on the population of pulsars that originate near the Galactic Center, we assume that each pulsar in our model is formed at a location within a few parsecs (well within the point spread function of HESS) around Sgr A$^*$. Once formed, however, each pulsar obtains a natal kick velocity which continuously carries it away from the Galactic Center, broadening the angular profile of the resulting gamma-ray emission. For simplicity, we adopt a uniform kick velocity of 400 km/s for each pulsar in our model, and an initial position of 1 parsec from Sgr A$^*$.

Of course not all pulsars originate near the Galactic Center, and much of the diffuse VHE emission observed from elsewhere in the sky could also originate from pulsars. Although we do not explore this possibility here, we consider it plausible that the diffuse TeV-scale emission observed from the Galactic Plane~\cite{Atkins:2005wu,Prodanovic:2006bq} could be generated by a population of such objects~\cite{Linden:2017blp}. Pulsars that do not originate in the innermost parsecs of the Galaxy could produce VHE emission that is significantly more spatially extended than that presented here.

\section{The Number of Pulsars Required to Generate the TeV Emission Observed From The Galactic Center}

In this section, we will use the spectrum reported by HESS, in conjunction with the measurements of Geminga by HAWC and Milagro, to estimate the number of pulsars located in the region surrounding the Galactic Center (generating the emission associated with both the central source and surrounding diffuse emission, as shown in Fig.~\ref{gammaspec}) . In carrying out this estimate, we implicitly assume that the pulsars near the Galactic Center deposit the same fraction of their spin-down power into electron-positron pairs as Geminga.

Drawing from the distribution of initial periods and magnetic field strengths described in Sec.~\ref{model}, we find that the average total spin-down power of the modelled pulsar population is $\dot{E} \approx 6.34 \times 10^{37}$\, erg$/$s $\times (R/1000)$, where $R$ is the birth rate of pulsars per Myr. Comparing this to the value for Geminga ($\dot{E}_{\rm Geminga} \approx 3.2\times 10^{34}$ erg/s) and correcting for the relative distances (we adopt $d_{\rm Geminga} = 250^{+230}_{-80}$ pc~\cite{Verbiest:2012kh} and $d_{\rm GC} =8250$ pc), we estimate that the total gamma-ray flux from the Galactic Center pulsar population should be equal to $(0.84-6.71) \times (R/1000)$ times that of Geminga. From HAWC's measurement of the VHE flux from Geminga, and after correcting for the higher energy densities in radiation and magnetic fields near the Galactic Center, this translates to a flux of $(3.25^{+8.72}_{-1.75})\times 10^{-12}$ TeV cm$^{-2}$ s$^{-1} \, \times (R/1000)$ at an energy of 7 TeV.

We compare this predicted flux to that reported by the HESS Collaboration. At 7 TeV, the flux observed by HESS from the combination of the central source and the surrounding annulus (as shown in the left and right frames of Fig.~\ref{gammaspec}) is $1.59^{+0.36}_{-0.34} \times 10^{-12}$ TeV cm$^{-2}$ s$^{-1}$. For an average birth rate of $R \simeq 490^{+580}_{-370}$ new pulsars per Myr, this gamma-ray flux can be fully accounted for by the VHE electrons and positrons injected from pulsars.

One should keep in mind that most of these pulsars will produce radio beams that are not aligned toward the Solar System, and will thus be impossible to detect with radio telescopes. For an estimated beaming fraction of 25\%~\cite{1998MNRAS.298..625T}, we predict the Galactic Center to contain between $\sim$25-190 pulsars younger than 1 Myr with radio beams oriented in our direction. Assuming that a typical pulsar remains radio bright for $\sim$10 Myr, this would imply that $\sim$250-1900 potentially observable radio pulsars should be present within 70 parsecs of the Galactic Center.\footnote{The uncertainty on the number of potentially observable pulsars near the Galactic Center is dominated in our calculation by the distance to Geminga. Future refinements of this quantity will enable us to more reliably predict the number of pulsars present.} For comparison, Ref.~\cite{Chennamangalam:2013zja} estimate that as many as $\sim$200 such sources could be present within the Galaxy's innermost parsec.

\section{The Angular Distribution of Very High Energy Emission From Pulsars Near the Galactic Center}

\begin{figure}
\includegraphics[width=3.0in,angle=0]{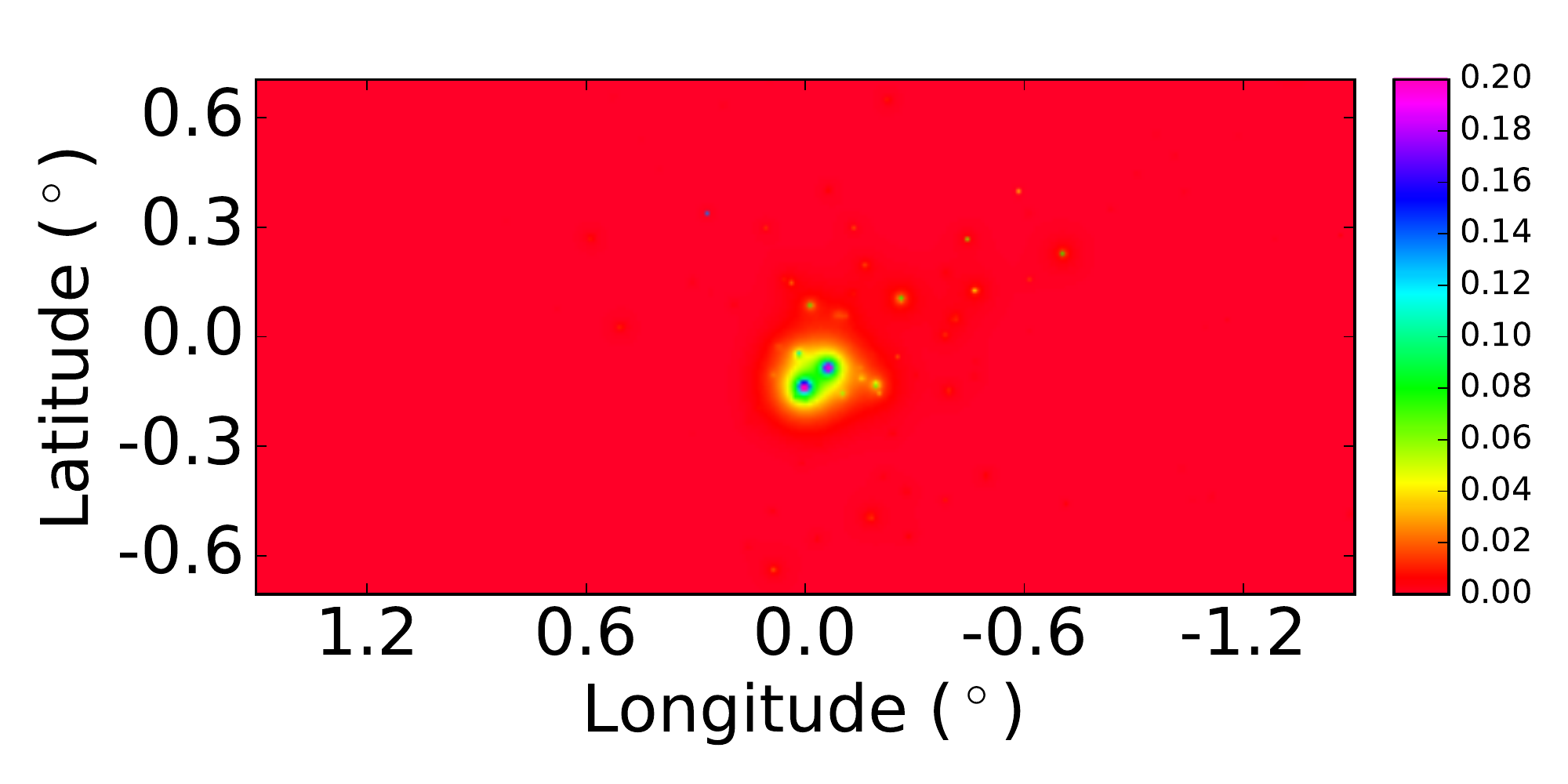}
\hspace{-0.4cm}
\includegraphics[width=3.0in,angle=0]{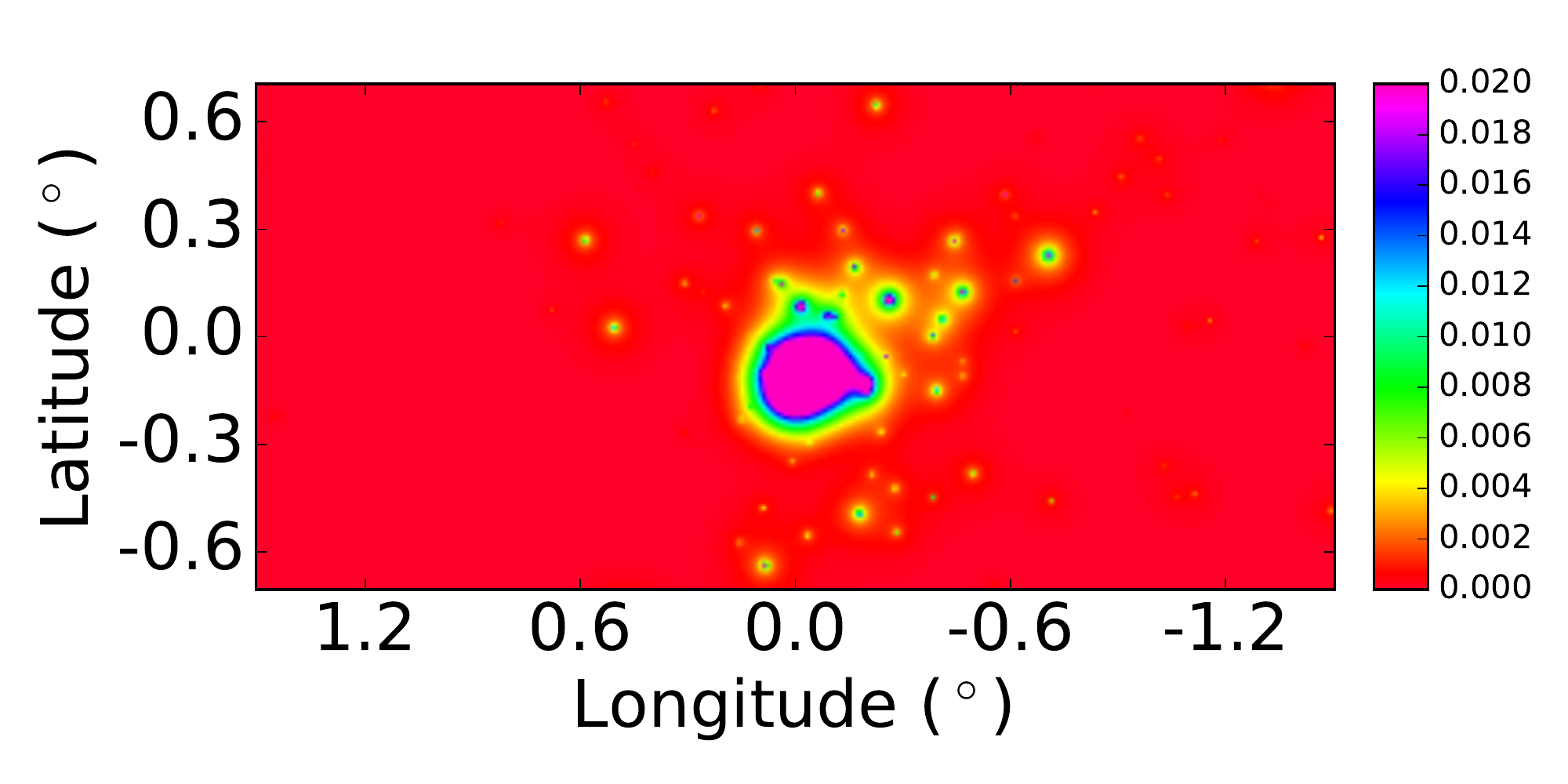} \\
\vspace{-0.6cm}
\includegraphics[width=3.0in,angle=0]{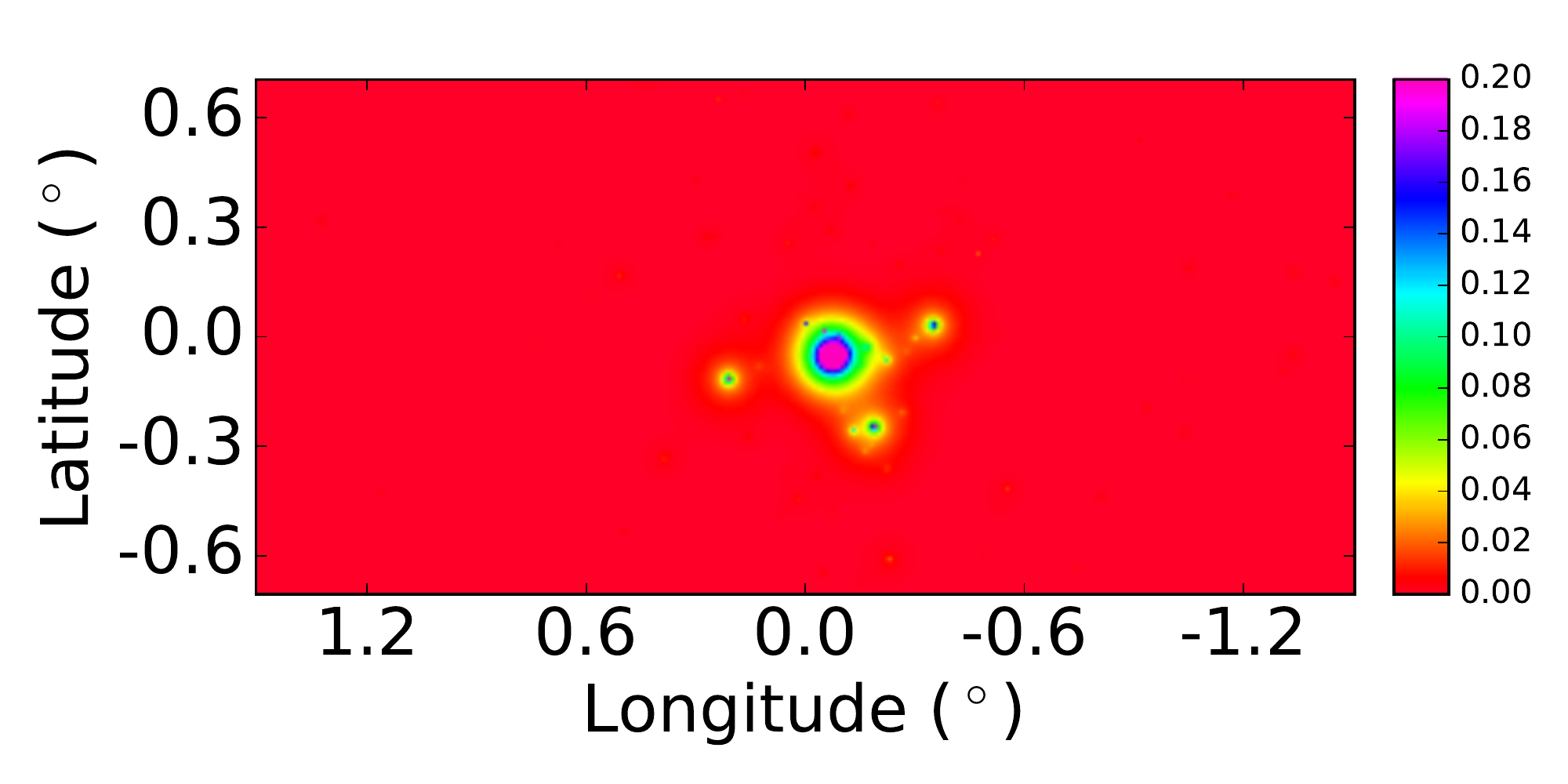}
\hspace{-0.4cm}
\includegraphics[width=3.0in,angle=0]{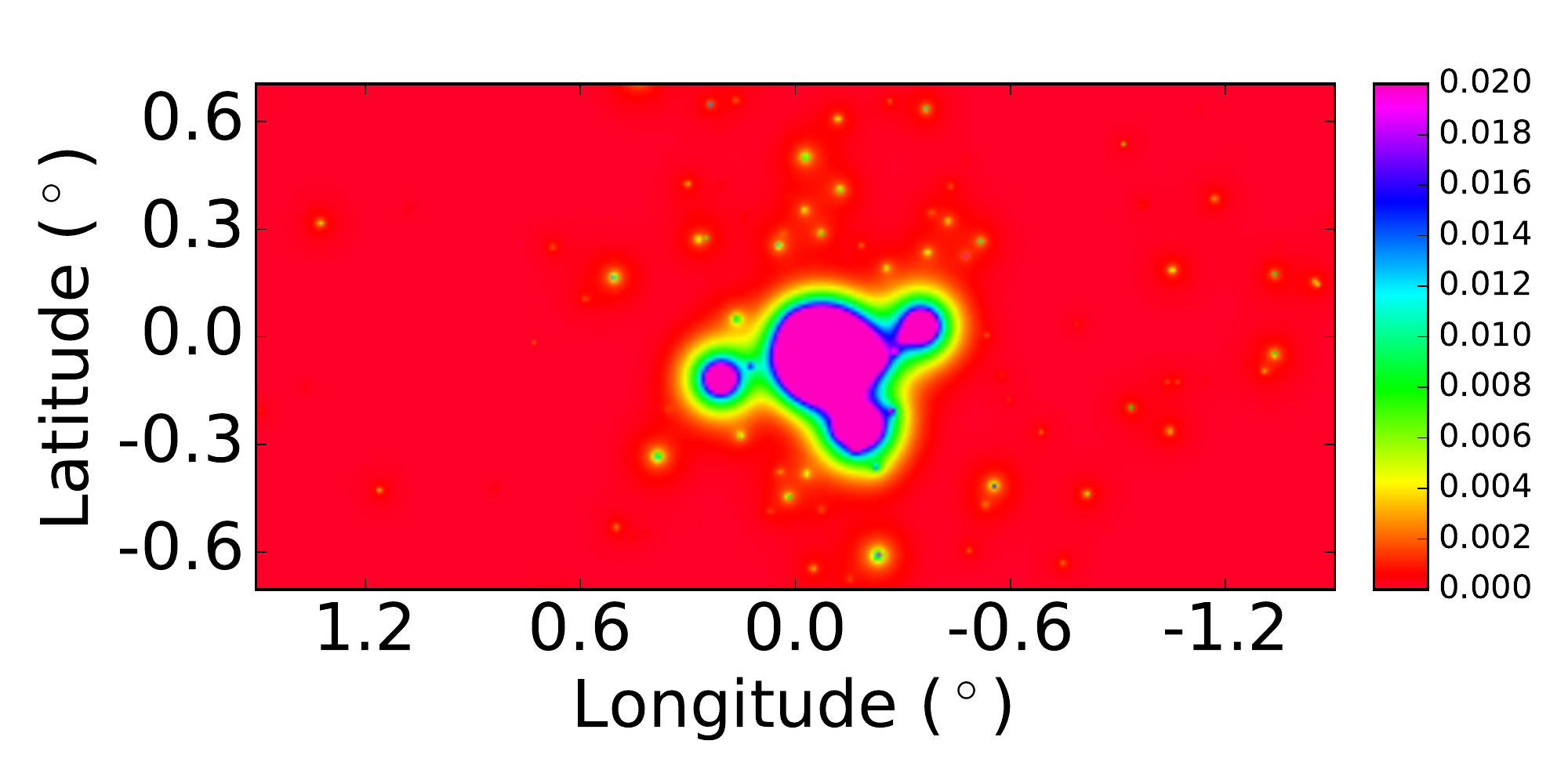} \\
\vspace{-0.6cm}
\includegraphics[width=3.0in,angle=0]{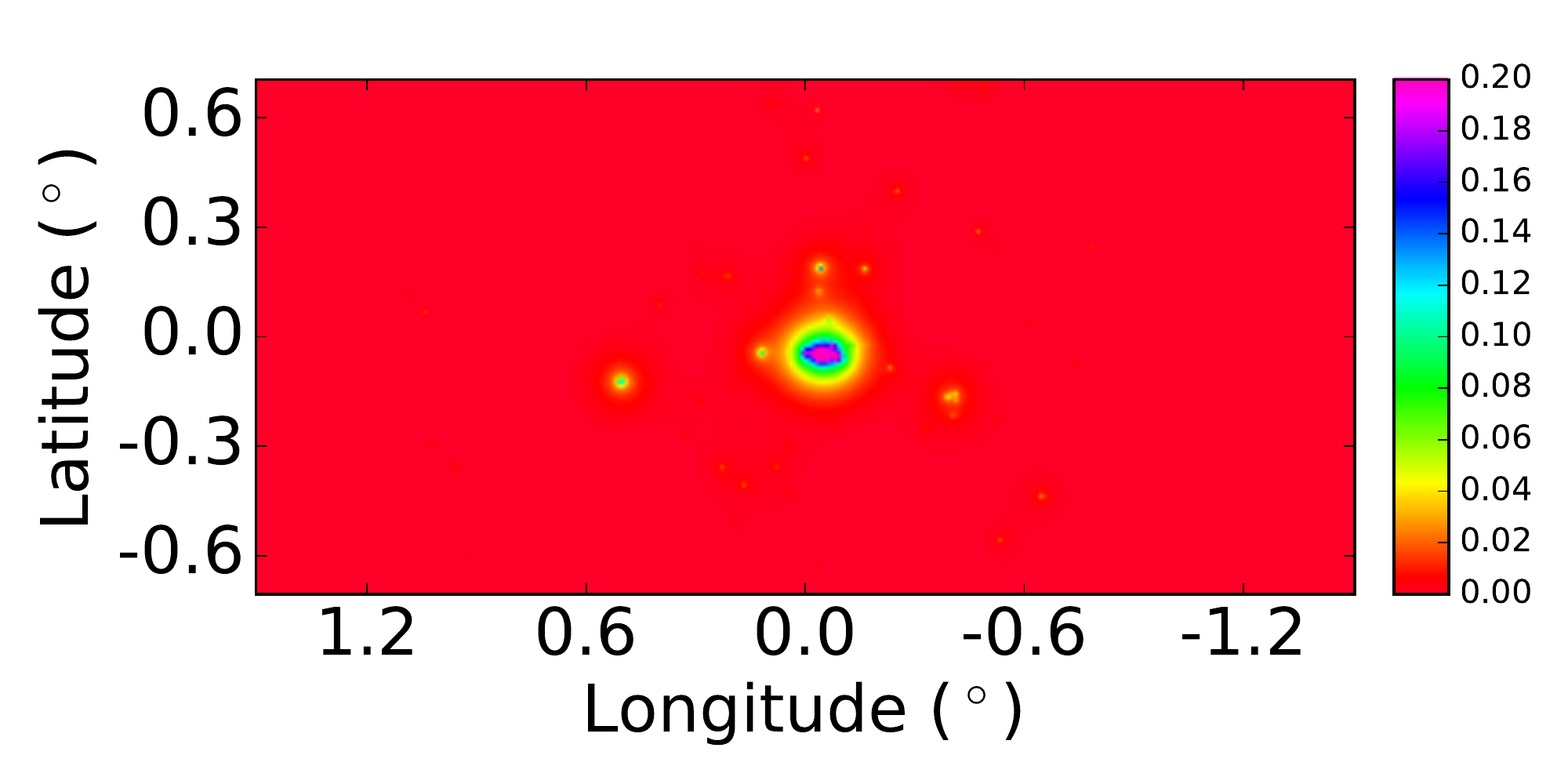}
\hspace{-0.4cm}
\includegraphics[width=3.0in,angle=0]{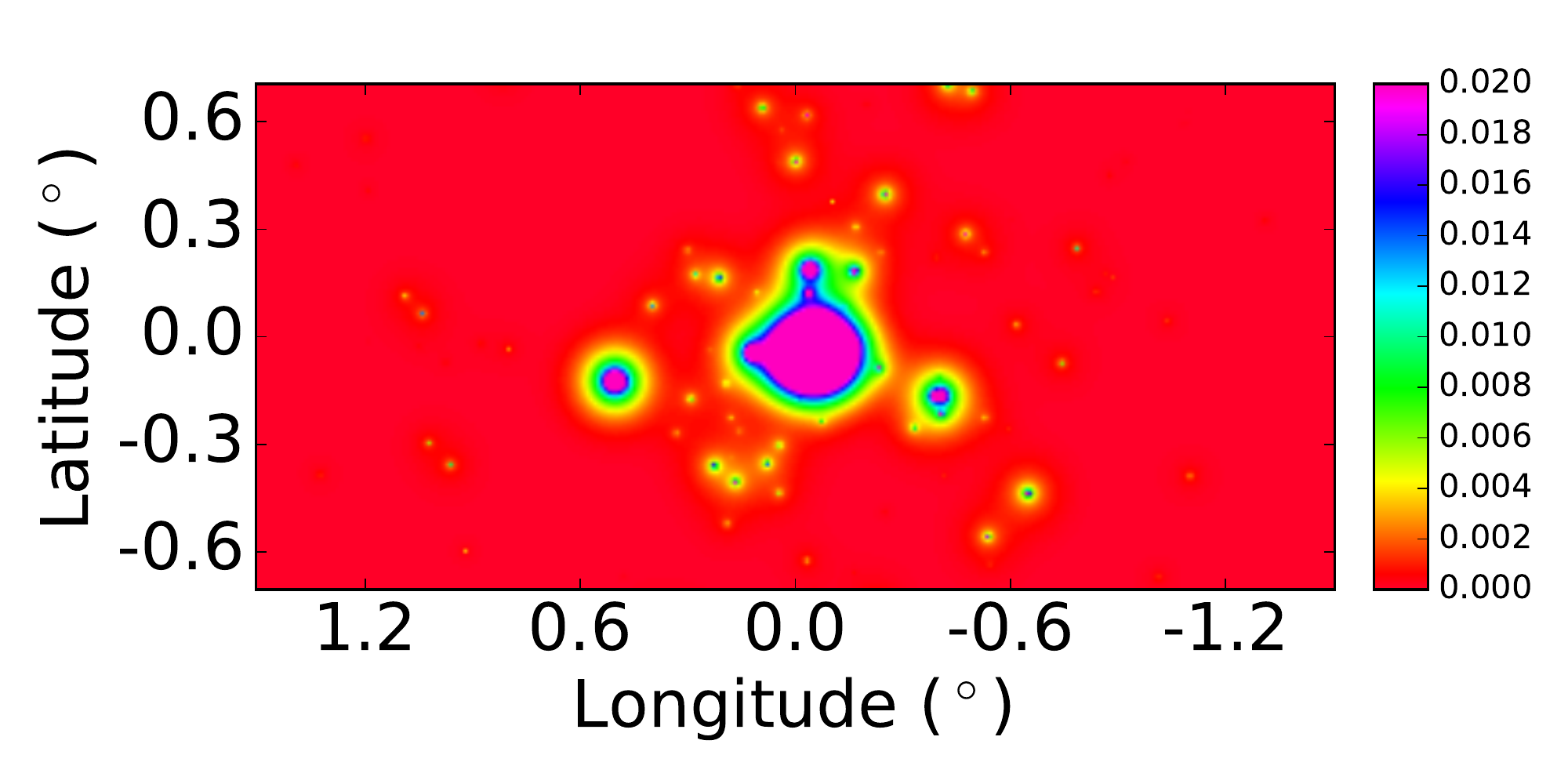} \\
\vspace{-0.6cm}
\includegraphics[width=3.0in,angle=0]{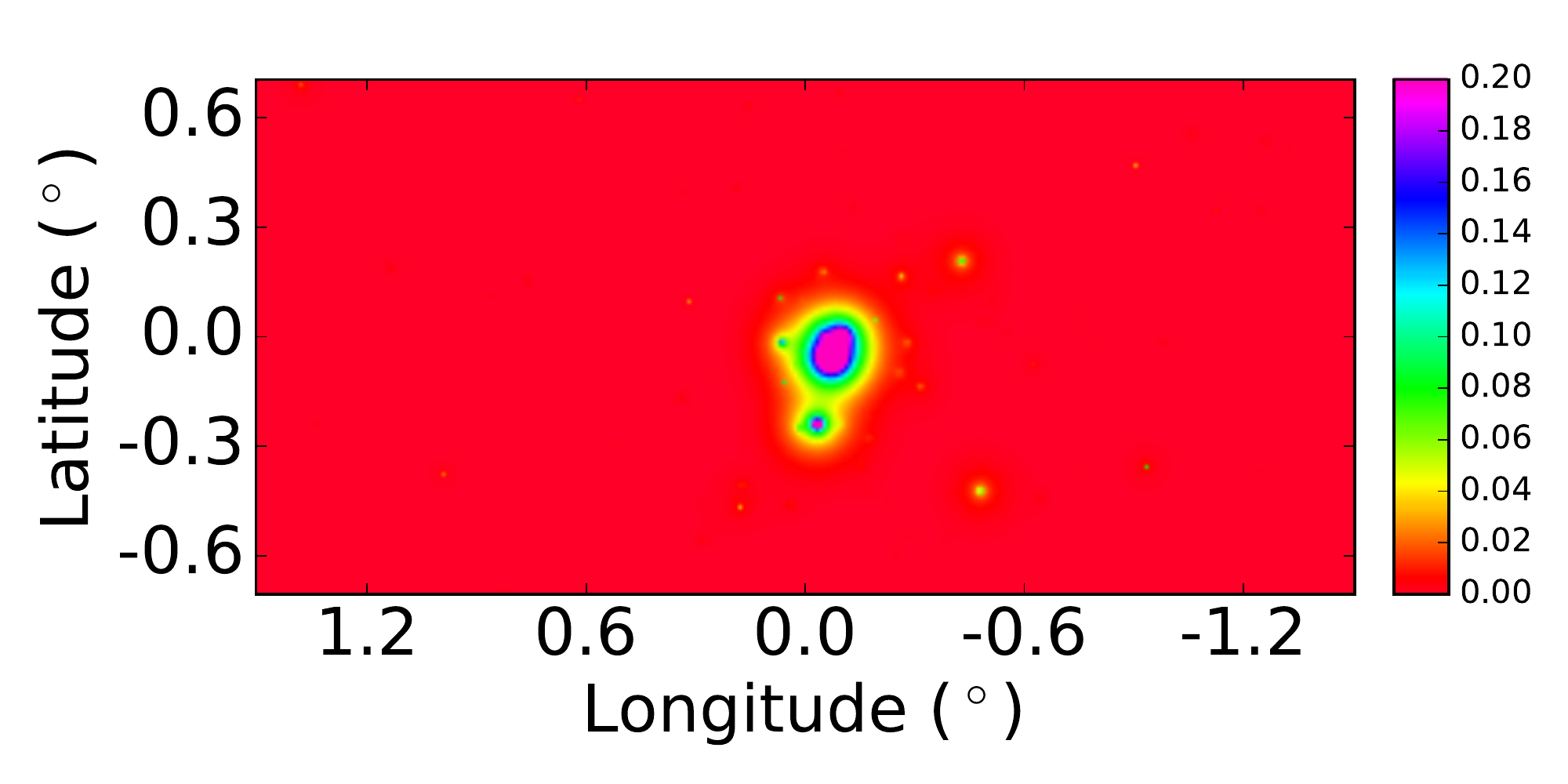}
\hspace{-0.4cm}
\includegraphics[width=3.0in,angle=0]{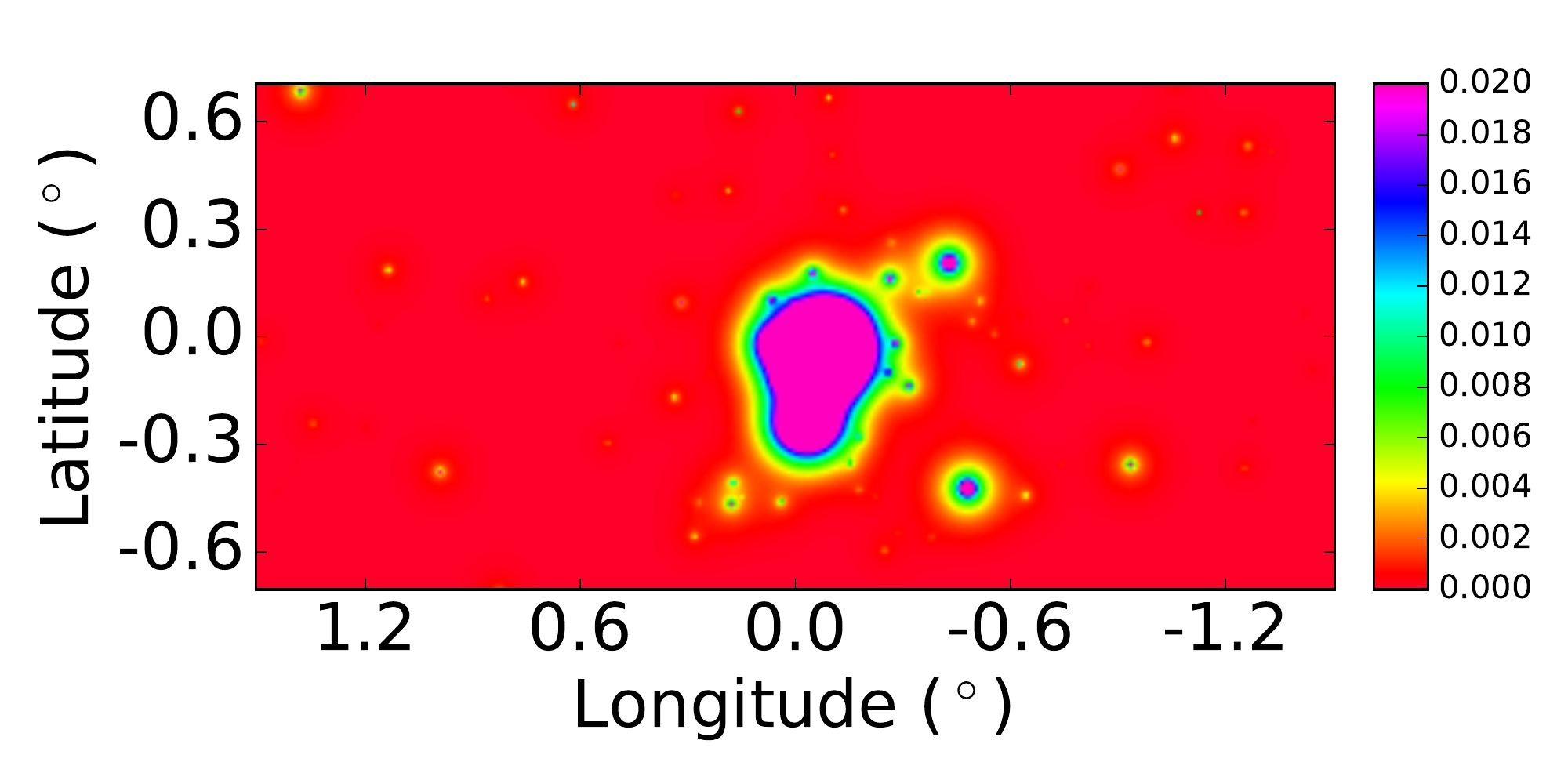} \\
\vspace{-0.6cm}
\includegraphics[width=3.0in,angle=0]{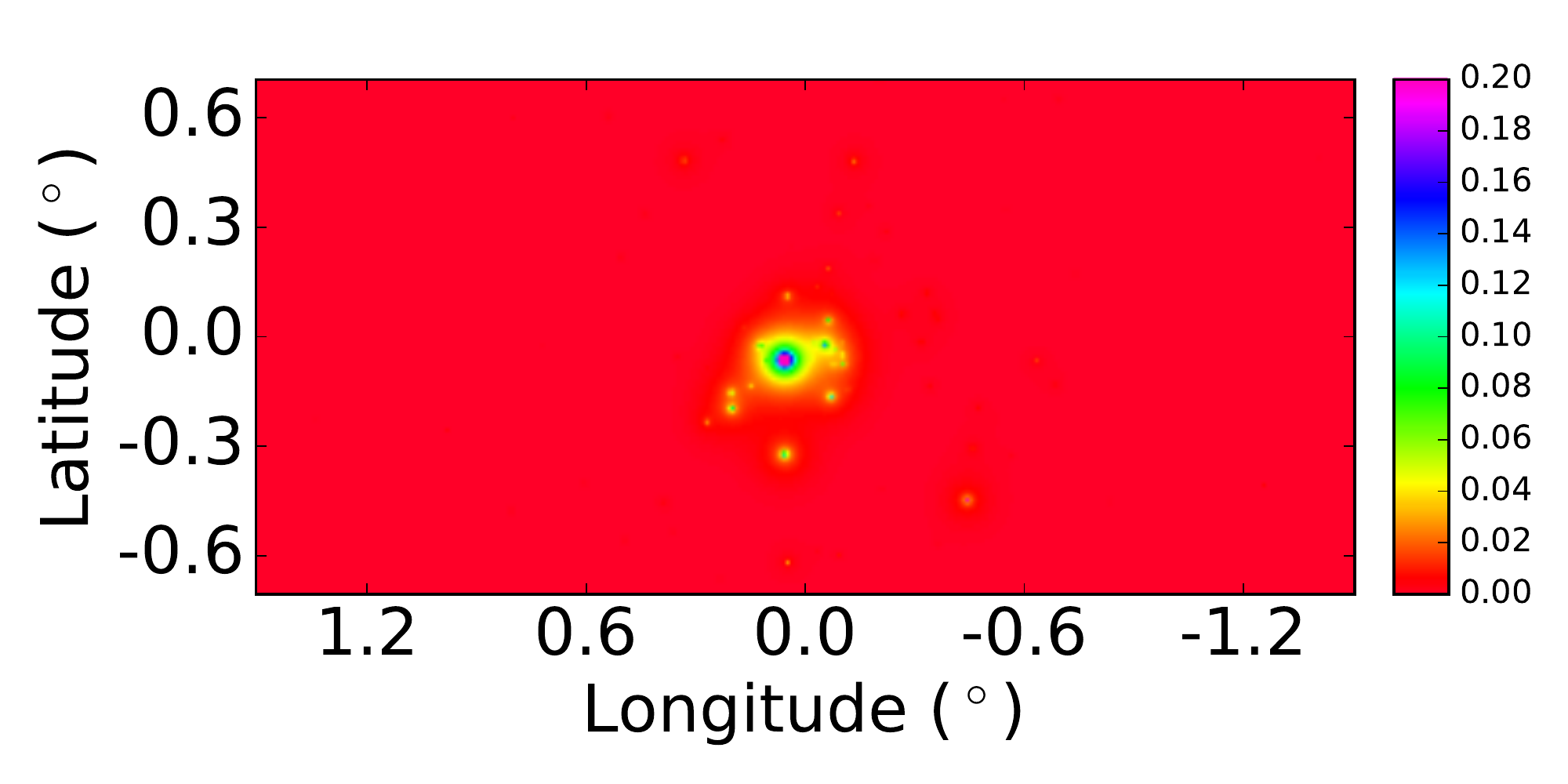}
\hspace{-0.4cm}
\includegraphics[width=3.0in,angle=0]{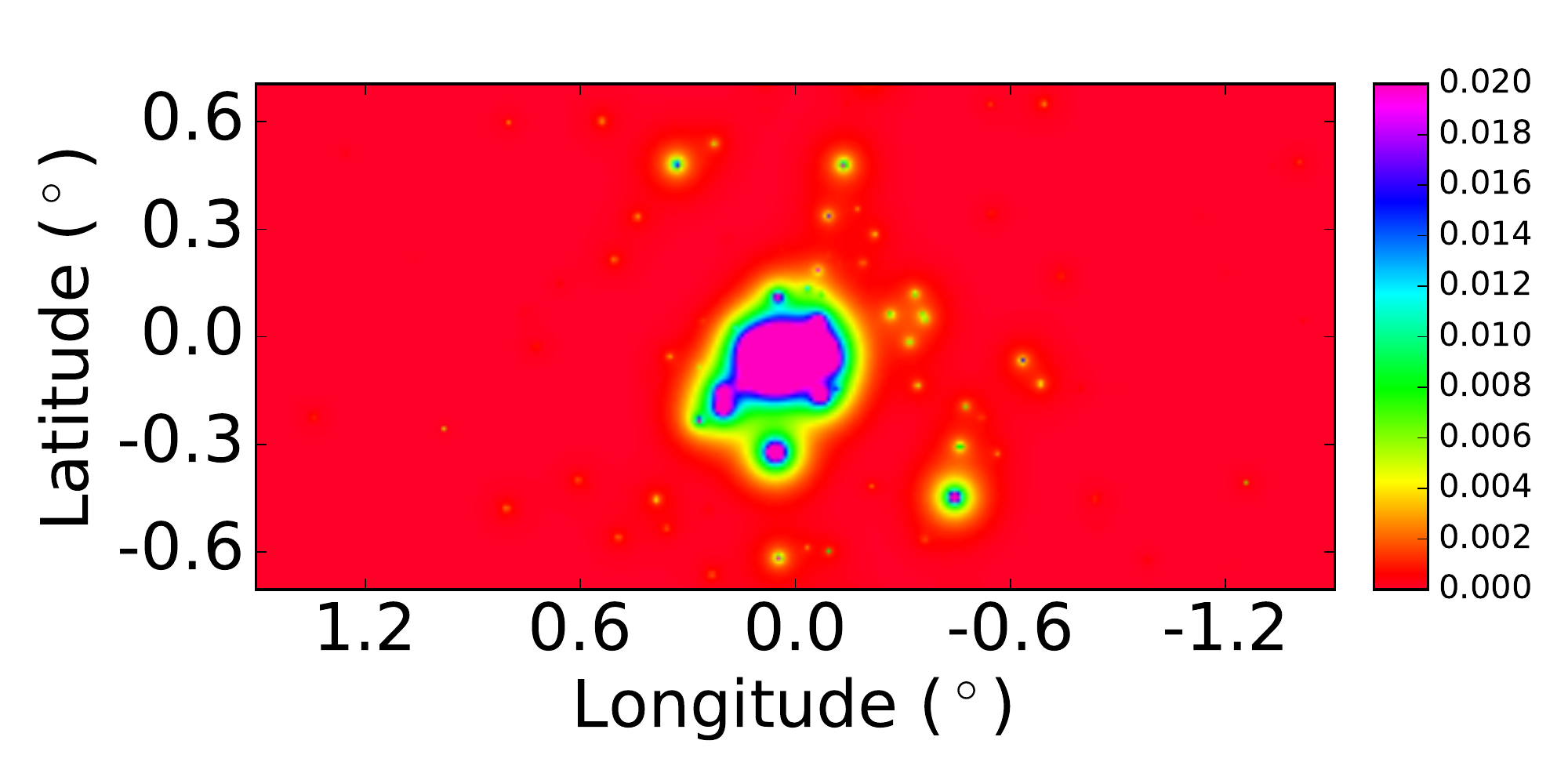}  \\
\vspace{-0.6cm}
\includegraphics[width=3.0in,angle=0]{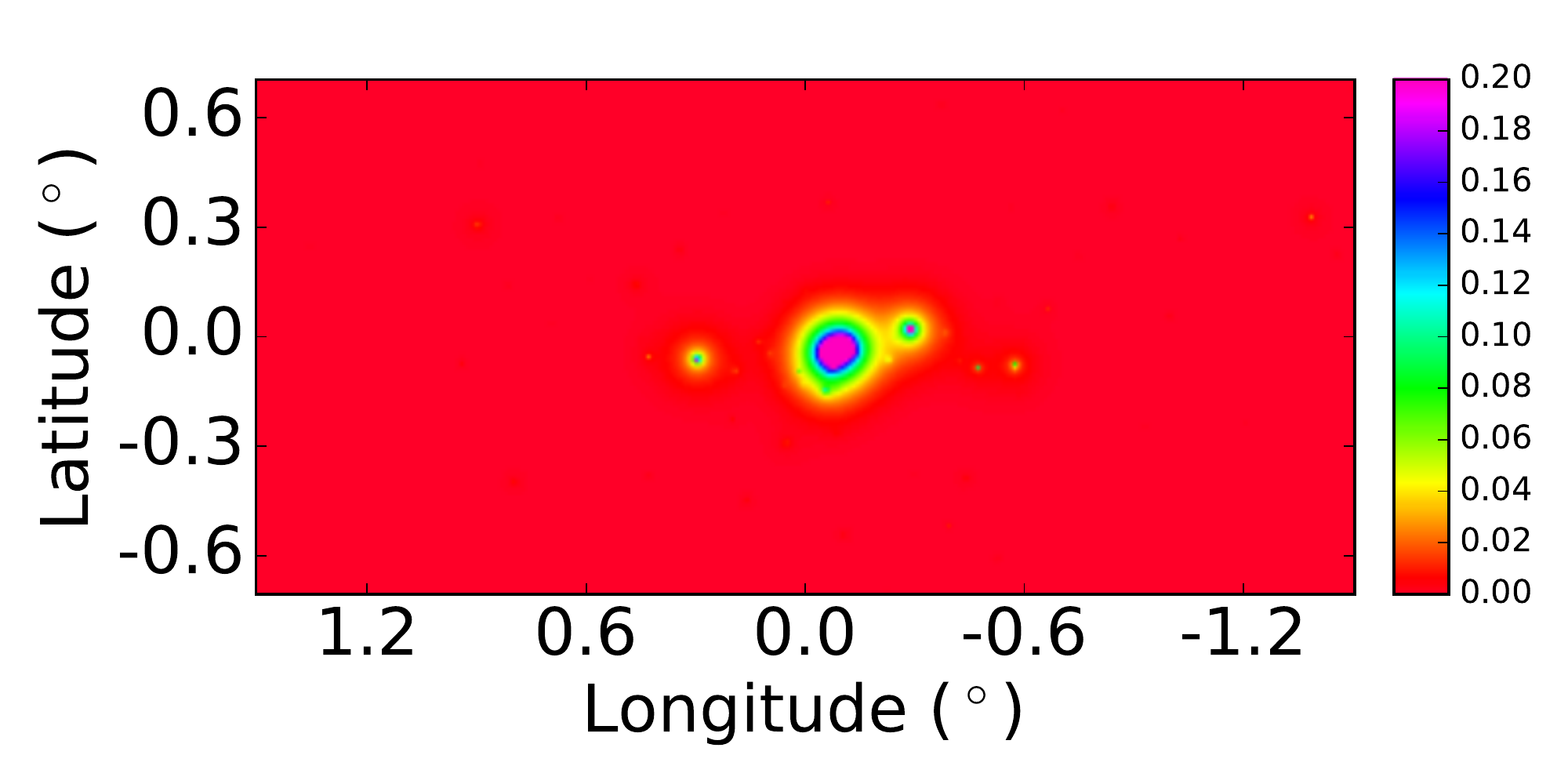}
\hspace{-0.4cm}
\includegraphics[width=3.0in,angle=0]{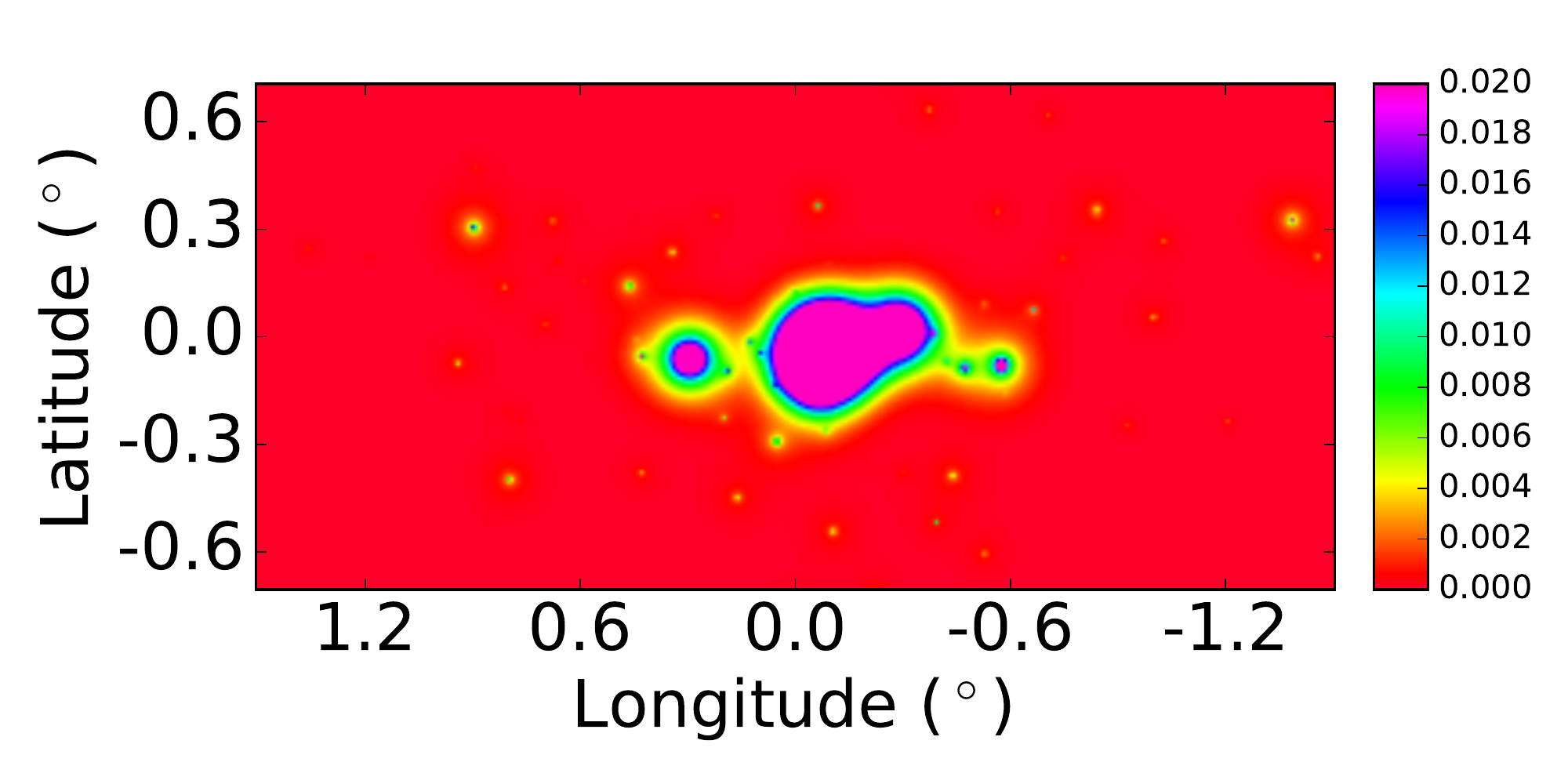}
%
\caption{The morphology of VHE emission from pulsars originating near the Galactic Center in our model, for six randomly chosen realizations, each with a neutron star birth rate of 200/Myr. The left and right frames correspond to the same maps, shown with different dynamical ranges.}
\label{morph}
\end{figure}

In Fig.~\ref{morph}, we plot the morphology of the VHE gamma-ray emission from pulsars originating near the Galactic Center for six randomly chosen realizations, each with a neutron star birth rate of 200 per Myr. In addition to calculating the trajectory of each simulated pulsar, we have assumed that the emission from each pulsar has a physical extent equal to that observed for Geminga (a Gaussian with a width of $2^{\circ} \times 250\, {\rm pc}/d\sim 0.06^{\circ}$) and have convolved the predicted emission by the HESS point spread function (which we approximate by a 0.06$^{\circ}$ Gaussian).\footnote{For clarification, it is a coincidence that the physical extent of a Geminga-like TeV halo and the point spread function of HESS are each described by a Gaussian of width 0.06$^{\circ}$.} We note that the physical extent of the TeV halos surrounding Galactic Center pulsars might differ substantially from those surrounding Geminga and B0656+14, owing to differences in the density of the interstellar medium, and in diffusion and energy loss processes. 

Broadly speaking, the main features of this simulated emission are consistent with those reported by the HESS collaboration, although a detailed comparison is made difficult by the fact that the predicted emission varies considerably depending on the brightest few pulsars that happen to be present at this particular point in time. That being said, from among the six randomly chosen realizations shown in Fig.~\ref{morph}, the ratio of the flux from within the HESS point spread function ($0.06^{\circ}$) around the Galactic Center to that from within the 0.2-0.5$^{\circ}$ (partial) annulus varies from between 0.3 and 1.1, consistent with the ratio of fluxes reported by HESS (as inferred by comparing the left and right frames of Fig.~\ref{gammaspec}).

We note that the ratio of the gamma-ray fluxes attributed to the central point source and to the surrounding annulus depends (to a similar degree) on four parameters: the velocity distribution of neutron star natal kicks, the evolution of the pulsar spin-down luminosity, the physical size of TeV halos, and the point-spread function of HESS. In each case, we have selected well-motivated values which were not fit to the HESS data. However, we stress that alterations in these parameters can affect the ratio of the central and diffuse fluxes.

\section{Millisecond Pulsars}

Thus far, we have focused on young pulsars, in contrast to recycled pulsars with millisecond-scale periods. The main reason for this is that, to date, no TeV-halos have been observed around any millisecond pulsars (MSPs), and thus we do not know what fraction of the spin-down power of these sources goes into the production of VHE electrons and positrons. It is widely believed, however, that MSPs are indeed likely to generate such emission~\cite{Venter:2015gga,Bednarek:2016gpp,Venter:2015oza}, as the modelling of their light curves favor the abundant production of multi-TeV electron-positron pairs~\cite{Venter:2015gga}.

Although no MSPs have been detected near the Galactic Center, the number of such objects present in the Inner Galaxy is not well constrained and could plausibly be large. The subject of MSPs in the Inner Galaxy has received a great deal of attention in recent years, as it has been argued~\cite{Abazajian:2010zy,Hooper:2010mq,Hooper:2011ti,Gordon:2013vta,Abazajian:2014fta,Cholis:2014lta,Petrovic:2014xra,Hooper:2015jlu,Brandt:2015ula,OLeary:2016cwz,Hooper:2016rap,Haggard:2017lyq,Ploeg:2017vai,Fermi-LAT:2017yoi} that a large population of such objects could plausibly be responsible for the Galactic Center gamma-ray excess observed by the Fermi Telescope~\cite{Goodenough:2009gk,Hooper:2010mq,Hooper:2011ti,Abazajian:2012pn,Hooper:2013rwa,Gordon:2013vta,Daylan:2014rsa,Calore:2014xka,TheFermi-LAT:2015kwa,TheFermi-LAT:2017vmf}.

The total luminosity of the Galactic Center gamma-ray excess is $L_{\gamma} \sim 2\times 10^{36}$ erg/s above 100 MeV, integrated within $0.5^{\circ}$ of the Galactic Center~\cite{Cholis:2014lta}. Given that the gamma-ray efficiency (averaged over $4\pi$ steradians) measured for the vast majority of MSPs observed by Fermi is between a few percent and unity~\cite{TheFermi-LAT:2013ssa}, this implies that the total spin-down power of this MSP population is required to be at least $\dot{E}_{\rm total} \sim (2-70) \times 10^{38}$ erg/s, which exceeds the total spin-down power of centrally located young pulsars by a factor of $\sim$10-500. Thus, while little is currently known about the VHE gamma-ray emission from MSPs, if these objects transfer more than a few percent of their total spin-down power into VHE pairs, the resulting inverse-Compton emission would exceed that observed by HESS from the Inner Galaxy. 

\section{Discussion and Summary}

The spectrum and morphology of the very high-energy (VHE) gamma rays observed from the Inner Galaxy have been interpreted as evidence that the Galactic Center (and Sgr A$^*$, in particular) accelerates protons up to $\sim$PeV energies, which then propagate outward and generate the observed emission through pion production. This scenario is further supported by the reported correlation between this emission and the distribution of molecular gas. 

In this article, we argue that the VHE emission from the Inner Galaxy is also likely to receive sizable contributions from pulsars, motivated by observations of nearby pulsars by HAWC and Milagro. In particular, HAWC's measurements of the spectrum and angular distribution of multi-TeV emission from Geminga and B0656+14 indicate that these sources deposit a significant fraction of their spin-down power into VHE electrons and positrons. If we assume that pulsars located in and around the Galactic Center also deposit a similar fraction of their energy into VHE pairs, then one can account for a large fraction of the VHE gamma-ray emission from the Inner Galaxy as observed by HESS and other ground-based telescopes.

The contribution of TeV halos to the VHE emission observed from the Inner Galaxy by HESS is well-motivated by three factors. First, the spectrum of the observed gamma-ray emission is consistent with that observed from the TeV halos surrounding the Geminga and B0656+14 pulsars. Second, the intensity of the observed emission is consistent with the expected pulsar population of the Galactic Center, based on the numbers of massive stars and low-mass X-Ray binaries observed in the region. In particular, based on a simple pulsar distribution model, we estimate that the observed emission requires a total birth rate of $490^{+580}_{-370}$ neutron stars per Myr, some $\sim$25-190 of which will constitute pulsars with radio beams directed toward the Solar System. Finally, the spatial extension of the observed TeV emission is consistent with a scenario where pulsars are transported out of the Galactic Center by neutron star natal kicks. While previous studies~\cite{Abramowski:2016mir,Fujita:2016yvk,Guo:2016zjl} have argued that leptonic gamma-ray models cannot produce the spatially extended emission observed from the Galactic Center, our model naturally avoids this constraint by transporting the sources of the mulit-TeV electrons outward and away from Sgr A$^{*}$.

In the years ahead, we expect a number of observations to refine and clarify this situation. Firstly, we anticipate that HAWC will measure the spectrum and angular extension of a significant number of pulsars~\cite{Linden:2017vvb}, allowing us to determine whether the emission observed from Geminga and B0656+14 is representative of the larger pulsar population. And although existing imaging atmospheric Cherenkov telescopes have not yet reported any significant detection of TeV-scale emission from Geminga or B0656+14~\cite{Ahnen:2016ujd,2015arXiv150904224A}, next generation telescopes (and in particular the Cherenkov Telescope Array) are expected to be sensitive to extended sources such as these. In the foreseeable future, we anticipate these VHE gamma-ray telescopes to accumulate a sizable catalog of both pulsars and pulsar wind nebulae~\cite{Linden:2017vvb,Abdalla:2017vci}. Future observations of the Inner Galaxy by the Cherenkov Telescope Array will also build upon and expand our knowledge of this region of the sky. Furthermore, during the same period of time, deep large-area radio surveys are anticipated to detect the first pulsars from the inner parsecs around the Galactic Center~\cite{OLeary:2016cwz,Chennamangalam:2013zja,Dexter:2013xga,Calore:2015bsx,Macquart:2015jfa,Bhakta:2017mpk}.

\bigskip
\bigskip
\bigskip

\textbf{Acknowledgments.} We would like to thank Alice Harding for valuable discussions. DH is supported by the US Department of Energy under contract DE-FG02-13ER41958. Fermilab is operated by Fermi Research Alliance, LLC, under contract DE- AC02-07CH11359 with the US Department of Energy. IC acknowledges support from NASA Grant NNX15AB18G and from the Simons Foundation. TL acknowledges support from NSF Grant PHY-1404311.

\bibliography{hawcig.bib}
\bibliographystyle{JHEP}

\end{document}